\documentclass[11pt]{article}
\usepackage{moriond,epsfig,amssymb}

\def\Journal#1#2#3#4{{#1} {\bf #2}, #3 (#4)}

\def\NPB{{\em Nucl. Phys.} B}
\def\PLB{{\em Phys. Lett.}  B}
\def\PRL{\em Phys. Rev. Lett.}
\def\PRD{{\em Phys. Rev.} D}

\def\JL{\em JETP Lett.}
\def\AJ{\em Astrophys. J.}
\def\AP{\em Astropart. Phys.}

\def\be{\begin{equation}}
\def\ee{\end{equation}}
\def\bea{\begin{eqnarray}}
\def\eea{\end{eqnarray}}

\begin{document}
\vspace*{4cm}
\title{SGOLDSTINOS: PRIMARIES OF ULTRA-HIGH ENERGY COSMIC RAYS}

\author{D.S. GORBUNOV}

\address{
Institute for Nuclear Research of the Russian Academy of
Sciences,\\
60th October Anniversary Prospect 7a, 117312, Moscow, Russia
}

\maketitle\abstracts{I describe supersymmetric extensions 
of the Standard Model with light
 {\it sgoldstinos} and discuss the explanation of  
Ultra High Energy Cosmic Rays   
above GZK cutoff in these models. Also I briefly discuss the possibility to
solve other cosmological and astrophysical puzzles, such as gamma-ray
bursts and dimming of high-redshift supernovae, within the models with light
sgoldstinos.}

\section{Introduction}
\label{sec:intro}
In {\it any model} with supersymetry being spontaneously broken exists
goldstino supermultiplet. In a set of models (pseudo)scalar
superpartners of goldstino ({\it sgoldstinos}) are light ($m<1$~GeV),
long-living and sufficiently strongly interacting. 

There are some puzzles in astrophysics and cosmology without relevant
solutions in the framework of the Standard Model. In particular,
Ultra High Energy Cosmic Rays (UHECRs) data require new physics and  
very few models
of new physics are consistent with UHECRs data; a phenomenologically
viable model, which is consistent, is a model with light
sgoldstino. Extremely light sgoldstinos might be responsible for
dimming of high-redshift supernovae. The emission of sgoldstinos from
supenovae might connect gamma ray bursts and supernovae explosions. 

In this talk I briefly discuss these statements.   

\section{Sgoldstino Couplings}
\label{sec:couplings}
In any supersymmetric extension of the Standard Model (SM) of particle
physics spontaneous supersymmetry breaking occurs due to 
non-zero vacuum expectation value of auxiliary component 
of some chiral or vector superfield. 
For definiteness, let us consider a model where 
chiral superfield
\[
\Phi=\phi+\sqrt{2}\theta\psi+\theta^2F_\phi
\]
obtains non-zero vacuum expectation value $F$ for the auxiliary
component. 

Then $\psi$ is a Goldstone fermion and 
interaction between goldstino and other fields are given by
Goldberger-Treiman formula
\[
{\cal L}_{int}=\frac{1}{F}J^{\mu}_{ SUSY}\partial_\mu\psi\;,
\] 
where $J^{\mu}_{SUSY}$ is a supercurrent. The superpartners of
goldstino,  
\[
\frac{1}{\sqrt{2}}(\phi+\phi^*)\equiv S\;,~~~~
\frac{1}{i\sqrt{2}}(\phi-\phi^*)\equiv P\;,
\]
are scalar and pseudoscalar {\it sgoldstinos}, respectively. 
Sgoldstinos remain massless at tree level and become massive due to corrections from high
order terms in K\"ahler potential. Provided these terms are
sufficiently suppressed sgoldstinos are light. The coupling constants between gauge bosons
and fermions of SM and sgoldstinos may be expressed in terms of $F$
and the soft supersymmetry breaking parameters of the 
Minimal Supersymmetric extension of the Standard Model (MSSM). 
Carrying out of these expressions
is straightforward. For instance, let us consider the vector
supermultiplet containing gauge boson $A_\mu$ and gaugino $\lambda$. 
In low-energy spectrum there is a soft term for gaugino mass 
${\cal L}_{soft}=M_\lambda\lambda\lambda+h.c.$ 
Since supersymmetry is broken {\it spontaneously}, this implies the
interaction between gauge superfield $W_\alpha$ and $\Phi$  
\begin{equation}
{\cal L}_{SUSY}=a\Phi W^\alpha W_\alpha\biggl|_{\theta^2}+h.c.
\label{gauge}
\end{equation}
with parameter $a$ obeying $aF=M_\lambda$. From Eq.~(\ref{gauge}) we
obtain the coupling constants for 
``sgoldstino-gauge boson-gauge boson'' vertices. In a similar way
one can derive the coupling constants between sgoldstinos and SM 
massive fermions: quarks and charged leptons (see Ref.~\cite{comphep} for
details).

\section{Phenomenology of Models with Light Sgoldstinos}
\label{sec:phenomenology}

Note, that all sgoldstino coupling constants mentioned above are
completely determined by soft terms of MSSM and parameter of
supersymmetry breaking $F$ but sgoldstino masses ($m_S,m_P$) remain 
free. Depending of the values of these two unknown parameters,
sgoldstinos may show up in different experiments. Phenomenologically
interesting models form four classes: 
\begin{enumerate}
\item Sgoldstino masses are of order of electroweak scale, 
$\sqrt{F}\sim1$~TeV --- sgoldstino may be produced 
in collisions of high energy particles at colliders~\cite{colliders}.

\item Sgoldstino masses $m_S,m_P\sim1$~MeV$\div1$~GeV, 
$\sqrt{F}\sim1$~TeV --- sgoldstinos may emerge in products of rare
decays of mesons~\cite{J,np}, such as $\Upsilon\to S(P)\gamma$, $J/\psi\to
S(P)\gamma$. 

\item Light sgoldstinos in models with flavor violation in the sector of
soft trilinear couplings, $A_{ij}\neq A\delta_{ij}$ --- sgoldstinos
may lead to flavor violating processes. Namely, they may contribute to 
FCNC (mass difference in the system of neutral mesons, CP-violation in
the system of neutral mesons)~\cite{fcnc,pr}, then, if 
kinematically allowed, they appear
in the product of rare decays, such as $t\to cS(P)$~\cite{pl}, $\mu\to eS(P)$,
$K\to\pi S$~\cite{np}, etc.

\item Sgoldstinos are lighter than 1~MeV --- these models may be tested
in low energy experiments~\cite{np}, such as reactor experiments, conversion in 
magnetic field, etc. Also sgoldstinos may play very important role in
astrophysics and cosmology~\cite{sn,nth,np}: 
they may distort cosmic microwave background spectrum, affect
supernovae explosion and cooling rate of stars, etc.  

\end{enumerate}

Direct independent measurement of MSSM soft supersymmetry breaking
terms and sgoldstino couplings provides the unique possibility to
estimate the scale of supersymmetry breaking $\sqrt{F}$.

\section{Sgoldstinos as UHECRs}
\label{sec:sgoldstino-primary}

Till now, there are no experimental evidence for sgoldstinos.  The
study of sgoldstino phenomenology results in obtaining bounds on its
coupling constants. Meanwhile there are some problems in astroparticle
physics and cosmology, which may be solved in model with sufficiently light
sgoldstinos.

We will concentrate here on the problem with Ultra High Energy Cosmic
Rays (UHECRs). Tens events with energy above Greisen-Zatsepin-Kuzmin
(GZK) cutoff~\cite{GZK}, $E>4\cdot10^{19}$~eV, suggest that something
appears to be missing
in our understanding of the sources, nature or propagation of UHECRs.

Indeed, the small-scale clustering of UHECR events suggests that the sources
are point-like on cosmological scales~\cite{Tinyakov:2001ic}. Recently,
a statistically significant correlation, at the level of chance
coincidence below $10^{-4}$, was found with the most powerful BL
Lacertae~\cite{Tinyakov:2001b} 
(the special type of active galaxies with jets pointed in our
direction).  The identified sources are at high redshift $z >0.1$, 
far exceeding the GZK distance of $R_{\rm GZK}\approx 50$~Mpc,
so that the primary ultra-high energy (UHE) particles can not be
protons.  The photon 
attenuation length for energies around $10^{20}$~eV is of order the
GZK cutoff distance, primarily due to the extragalactic radio
backgrounds, so one can conclude that
UHECRs with energies around $10^{20}$~eV are very unlikely to be
photons. The primaries of UHECRs should be some particles traveling 
for cosmological distances unattenuated (without significant energy loss).  
The only Standard-Model particles which can reach our Galaxy without
significant loss of energy are neutrinos. In the framework of Standard Model 
neutrinos
produce nucleons and photons via resonant $Z$-production~\cite{weiler} 
with relic
neutrinos clustered within about 50~Mpc from the Earth, giving rise to
angular correlations with high-redshift sources. 
However, for the production rates to be sufficiently high, this
scenario requires enormous neutrino fluxes and an extreme clustering
of relic neutrinos with masses in the (sub-)eV range~\cite{zburst}.  

Thus UHECRs with energies around $10^{20}$~eV may point at the new
physics beyond the Standard Model. Light sgoldstinos might be such a
new physics.

Indeed, the GZK cutoff can be avoided also if the UHECRs consist of certain
new particles, 
which can traverse the universe unimpeded at high energies. 
Such particles must fulfill several requirements to be candidates for
UHECRs:
\begin{itemize}
\item  they must live long enough to reach us from cosmological
distances;  
\item they must not lose too much energy in interactions with the
CMBR and other background radiations or in extragalactic magnetic
fields;  
\item they must interact sufficiently strongly in or near our
Galaxy or in the Earth's atmosphere to produce the observed UHE
events; 
\item their interactions must allow for the production of a
significant flux at the source. 
\end{itemize}

All these requirements may be fulfilled in the models with light
sgoldstinos.~\cite{semikoz}. For light scalar
sgoldstino~\footnote{Light pseudoscalar sgoldstino may be considered
in the same way.} the relevant interactions are
$$
{\cal L}=\frac{M_{\lambda_3}}{2\sqrt{2}F} S G_{\mu\nu}^aG^{\mu\nu}_a\;,~~~~~~
{\cal L}=\frac{M_{\gamma\gamma}}{2\sqrt{2}F}S F_{\mu\nu}F^{\mu\nu}\;,
$$
where 
$$
M_{\gamma\gamma}=M_{\lambda_1}\cos^2\theta_W+M_{\lambda_2}\sin^2\theta_W
$$
and $M_{\lambda_i}$ are MSSM gaugino soft mass parameters. 
If $M_S<200$~MeV, the dominant decay mode is into two
photons:
$$
\Gamma(S\to\gamma\gamma)={M_{\gamma\gamma}^2M_S^3\over32\pi F^2}\;,
$$
because the direct coupling to electrons is suppressed by
electron mass. 
This light particle with the energy $E_S$ propagates through the
Universe without decay if
$$
R_{\rm Universe}\lesssim L_{\rm decay} = 
{E_S\over \Gamma(S\to\gamma\gamma)M_S}\;.
$$
Therefore, if this particle is supposed to reach us from cosmological 
distances one has to require
\begin{equation}
\sqrt{F}\gtrsim 1.5\times10^6~{\rm GeV}
\left({10^{20}~{\rm eV}\over E_S}\right)^{1/4}
{M_S\over 10~{\rm MeV}}~~~~~~~{\rm at}~~~M_{\gamma\gamma}=100~{\rm GeV}\;.
\label{S-lifetime-gamma}
\end{equation}
The tiny photon coupling
required by Eq.~(\ref{S-lifetime-gamma}) guarantees also the absence of a
GZK cutoff for sgoldstino: the interactions with photon background and
(extra)galactic magnetic field are negligible.   

Both the production of sgoldstino at the source and their
interaction in the atmosphere require rather large cross sections,
comparable to strong ones.  
Typical energies of UHECR interactions with nucleons in c.m. frame are 
$E_{\rm cm}\approx 100\div300$~TeV.  The interaction cross section
with nucleons at such energies may be estimated as 
$$
\sigma_S = \sigma_{strong}\cdot\frac{\alpha_S}{\alpha_s}\;.
$$ 
The suppression factor
\begin{equation}
 \frac{\alpha_S}{\alpha_s} = \frac{(E_{\rm cm}M_{\lambda_3})^2}{32\pi F^2\alpha_s}  
\label{suppress}
\end{equation}
should not be very small. Since our particle exhibits strong
interactions 
one can estimate sgoldstino mean free path $\ell_S$ in the Earth's
atmosphere by analogy with the proton mean free path $\ell_p$ as
$$
\ell_S=\ell_p\cdot\frac{\alpha_{strong}}{\alpha_S}\;.
$$
To initiate an atmospheric
shower, $S$ should have a relatively small mean free path.  Assuming
$\ell_S<10\,\ell_p$ and using Eq.~(\ref{suppress}) and $\alpha_s=0.1$
one obtains 
\begin{equation}
\sqrt{F}\lesssim
1.3\times 10^4~{\rm GeV}
\left({E_S\over10^{20}~{\rm eV}}\right)^{1/4}~~~~~~~{\rm at}~M_{\lambda_3}=500~{\rm GeV}\;.
\label{g-S-atmosphere}
\end{equation}
The inequalities~(\ref{S-lifetime-gamma}) and~(\ref{g-S-atmosphere})
determine the region in the space of parameters $M_S$, $\sqrt{F}$,
which is suitable for the explanation of the UHECRs above the
GZK cutoff. Strong coupling to gluons guarantees that sgoldstino will be effectively produced in the 
high-energy tail of the proton spectra by
proton-proton collisions while their production at low energies will
be negligible.  Therefore, we can expect that the proton flux from the
source at low energies will continue with the same slope at high
energies due to sgoldstino component. Only part of the initial proton
energy will be transfered to energetic sgoldstinos; most of them will be
produced on the peak of the gluon distribution function with $E
\approx 0.1 E_p$.  
However, once produced they will escape more easily
from the source compared to protons precisely because their cross
section is smaller.

\begin{figure}
\epsfig{file=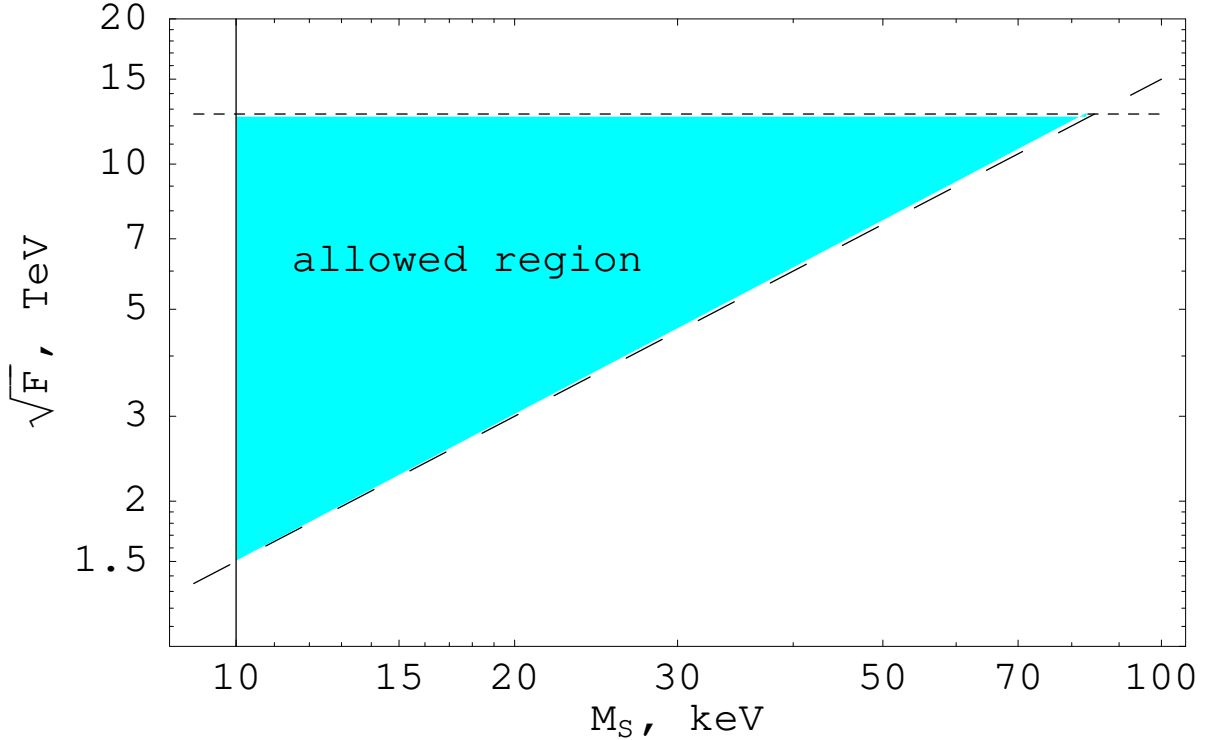, width=1.0\textwidth}
\caption{Allowed region for the parameters
  \protect$(M_S,\sqrt{F})$ at $M_{\gamma\gamma}=100$~GeV and gluino
mass $M_{\lambda_3}=500$~GeV. The short-dashed line corresponds to the
  limit~(\protect\ref{g-S-atmosphere}), the long-dashed line
  to~(\ref{S-lifetime-gamma}).  Sgoldstinos with masses less than
  10~keV (vertical solid line) are ruled out~\protect\cite{np} 
by the helium-burning life-time of horizontal-branch stars.}
\label{fig:ex2}
\end{figure}

A variety of experimental limits on models with light sgoldstinos has
been derived in Ref.~\cite{np}. These limits significantly constrain
the regions of parameter space of the phenomenologically viable models
with light sgoldstinos. In Fig.~\ref{fig:ex2} we
present the region of parameter space where sgoldstinos may act as
primaries of UHECRs and are not excluded by other limits.

\section{Discussions}
\label{sec:discussions}
The allowed region in Fig.~\ref{fig:ex2} suggests that the
supersymmetry breaking scale should be very low, 
$\sqrt{F} \sim 1\div10$~TeV. Hence our
light sgoldstino model can be tested in searches for rare decays of
$J/\psi$ and $\Upsilon$ and in reactor experiments (for details see
Ref.~\cite{np}).  This low scale of supersymmetry
breaking may be also tested at new generation accelerators like
upgraded Tevatron and LHC.  Note, that sgoldstino contributions 
to FCNC and lepton 
flavor violation are strong enough~\cite{np} 
to probe the supersymmetry breaking
scale up to $\sqrt{F}\sim10^4$~TeV if
off-diagonal entries in squark (slepton) mass matrices are close to
the current limits in the MSSM. Thus our light-sgoldstino scenario for
UHECRs allows only small flavor violation in the scalar sector of
superpartners.
 
Finally, let me note that sgoldstino might be responsible for other
astrophysical and cosmological puzzles. 
For instance, light sgoldstinos emitted from
supernovae might be a potential explanation for the origin of the
gamma-ray bursts.~\cite{baksan} Extremely light pseudoscalar
sgoldstino ($M_P\sim10^{16}$~eV) might be responsible for dimming of
high-redshift supernovae: emitted photons might convert into
sgoldstinos on extragalactic magnetic field. This solution does not
require dark energy component, in particular, non-zero cosmological
constant. In Ref.~\cite{dimming} the suitable region of parameter
space of the model with light axion ($f_a,m_a$) was found. The
suitable region of parameter space for sgoldstino solution may be
found by a simple mapping: $f_a\to F/M_{\gamma\gamma}$, $m_a\to M_P$.    

\section*{Acknowledgments}
It is a pleasure to thank Viatcheslav Ilyin, Barbara Mele, Georg
Raffelt, Valery Rubakov, Andrei Semenov and Dmitry Semikoz for fruitful
collaborations. 


\end{document}